\documentclass{mem}
\usepackage{natbib}\usepackage{txfonts}\usepackage{balance}
\usepackage{graphicx}
\usepackage[a4paper,breaklinks,dvipdfm]{hyperref}
\idline{75}{282}
\begin{document}
\def\teff{$T\rm_{eff }$}
\def\kms{$\mathrm {km s}^{-1}$}

\title{Photometric, astrometric, and spectroscopic survey of the old
  open cluster Praesepe}

   \subtitle{}

\author{
S. \,Boudreault\inst{1,2} 
\and N. \, Lodieu\inst{1,2}
          }

  \offprints{S. Boudreault}
 
  \institute{Instituto de Astrof\'{i}sica de Canarias (IAC), C/V\'{i}a
    L\'{a}ctea s/n, E-38200 La Laguna, Tenerife, Spain \and
    Departamento de Astrof\'{i}sica, Universidad de La Laguna (ULL), E-38206 La Laguna, Tenerife, Spain\\
    \email{szb@iac.es,nlodieu@iac.es} }

\authorrunning{Boudreault \& Lodieu}

\titlerunning{Survey of the old open cluster Praesepe}

\abstract{We analysed the wide-field near-infrared survey of the
  Praesepe cluster carried out by the UKIRT Infrared Deep Sky Survey
  (UKIDSS) Galactic Clusters Survey (GCS) and released by the Data
  Release 9 (DR9). We compare our Praesepe mass function (MF) with the
  ones of the Pleiades, $\alpha$ Per, and the Hyades. We also present
  preliminary results of a spectroscopic follow-up for the low mass
  members ($M$$\leq$0.1\,M$_\odot$) in Praesepe, $\alpha$~Per and
  Pleiades using the Optical System for Imaging and low Resolution
  Integrated Spectroscopy (OSIRIS) mounted on the 10.4m Gran
  Telescopio Canarias (GTC). We also present the optical spectrum of
  the first L dwarf in Praesepe.}

\maketitle{}

\section{Introduction}

Over the past decades, open clusters have been the subject of many
studies \citep[e.g.][and references therein]{bastian2010}.  Such
studies have brought new insights into brown dwarf (BD) formation, on
the discovery of young L and T dwarfs and free-floating planets and on
our understanding of the stellar/substellar MF and their populations
in the Galactic field and in open clusters. The extension of MF
studies to older clusters is vital as it allows us to study the
intrinsic evolution of BDs and how the stellar and substellar
population itself evolves. Praesepe is an interesting open cluster to
study the MF in the stellar and substellar regimes, considering its
age \citep[590$^{+150}_{-120}$\,Myr;][]{fossati2008}, distance
\citep[181.97$^{+5.96}_{-5.77}$\,pc;][]{vanleeuwen2009},
proper motion \citep[$\mu_\alpha\cos{\delta}$\,=\,$-$35.81$\pm$0.29
mas/yr and $\mu_\delta$\,=\,$-$12.85$\pm$0.24
mas/yr;][]{vanleeuwen2009}, and the low extinction towards this
cluster \citep[$E(B-V)$\,=\,0.027$\pm$0.004\,mag;][]{taylor2006}.

\section{Astrometric and photometric survey of Praesepe}

\begin{figure}[t!]
\resizebox{\hsize}{!}{\includegraphics[clip=true]{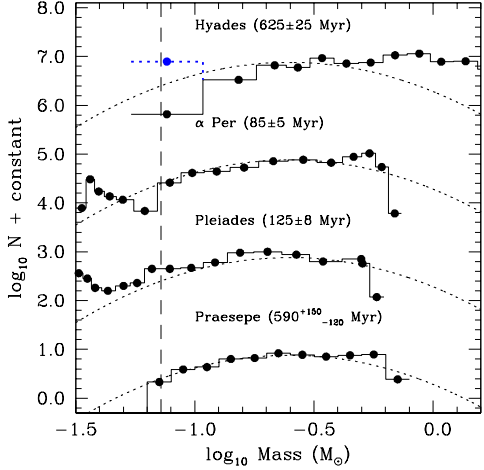}}
\caption{\footnotesize MFs of the Hyades, $\alpha$ Per, the Pleiades,
  and Praesepe. The blue dotted mass bin in the Hyades MF includes the
  12 L dwarf candidates from \citet{hogan2008}.  We also show
  the system Galactic field star MF \citep{chabrier2005} as the
  dotted curved lines and the substellar limit as a vertical dashed
  line. We normalised all the MFs to the log-normal fit of
  \citet{chabrier2005} at $\sim$0.3\,M$_\odot$ (log$M$$\sim$$~$0.5).}
\label{mf-all}
\end{figure}

We analysed the wide-field ($\sim$36 square degrees) near-infrared
($ZYJHK$) survey of the Praesepe cluster carried out by the DR9 of the
UKIDSS GCS\@. We selected 1,116 candidate cluster members of Praesepe
based on astrometry and five-band photometry. With our candidate list,
we derived the MF of Praesepe from 0.6 down to 0.072\,M$_\odot$. We
observed that our determination of the MF of Praesepe differ from
previous studies: while previous MFs present an increase from 0.6 to
0.1\,M$_\odot$, our MF presents a decrease \citep{boudreault2012}.  In
Fig.\ \ref{mf-all} we compare our Praesepe MF with the ones of the
Pleiades \citep{lodieu2012a}, $\alpha$~Per \citep{lodieu2012b}, and
the Hyades \citep{bouvier2008}. We concluded that our MF of Praesepe
is more similar to the MFs of $\alpha$~Per and the Pleiades, although
they are respectively of 85$\pm$5\,Myr \citep{dbyn2004} and
120$\pm$8\,Myr \citep{stauffer1998}, compare to 625$\pm$50\,Myr for
the Hyades \citep{bouvier2008}.

% FIGURE 1
%

\begin{figure}[t!]
\resizebox{\hsize}{!}{\includegraphics[clip=true]{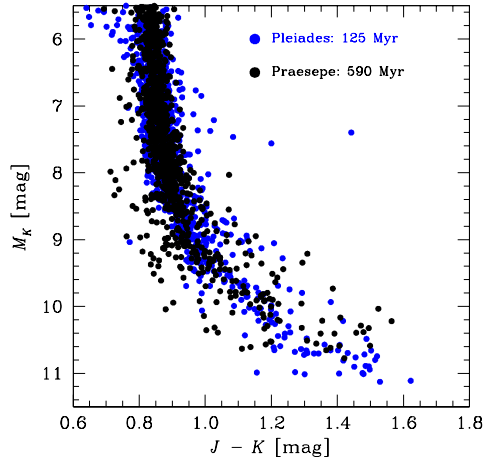}}
\caption{\footnotesize ($J-K$,$M_K$) diagram for the Pleiades (blue) and
  Praesepe (black).}
\label{cmd-plei-prae}
\end{figure}

One possible explanation for the discrepancy is an incompleteness of
the MF of the Hyades from \citet{bouvier2008}.  We noticed that adding
12 L dwarf cluster candidates in the Hyades \citep{hogan2008} reduces
the discrepancy between the two MFs (Fig.\,\ref{mf-all}).  Another
possible explanation for the discrepancy is an overestimation of the
age of Praesepe.  The fact that the MF of Praesepe is more similar to
the ones of $\alpha$~Per and the Pleiades than the one of the Hyades,
implies a similar dynamical evolution history. In addition, the
($J-K$,$M_K$) CMD of Praesepe and the Pleiades shows an overlap
between the two sequences, pointing towards a similar age/distance
combination for both regions (Fig.\ \ref{cmd-plei-prae}).

% FIGURE 2
%

\section{Spectroscopic survey}

We are embarked in a spectroscopic follow-up of low-mass member
candidates ($\leq$0.1\,M$_\odot$) selected in $\alpha$~Per, the Pleiades,
and Praesepe to constrain their membership, using OSIRIS
on the 10.4\,m GTC telescope in the Roque de Los
Muchachos Observatory in La Palma (Canary Islands). We used the R300R
grism and a 1.0\,arcsec slit with a 2$\times$2 binning, yielding a
spectral resolution of R\,=\,348 at 6865\AA{}.

\begin{figure}[t!]
\resizebox{\hsize}{!}{\includegraphics[clip=true]{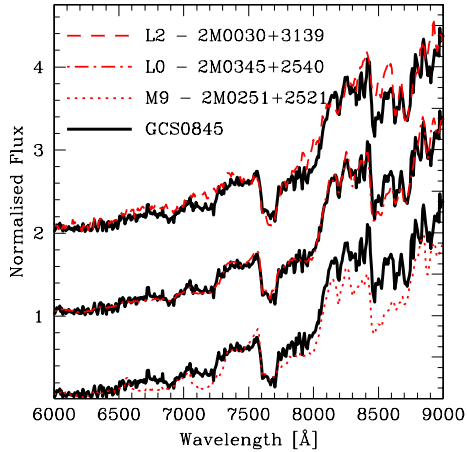}}
\caption{\footnotesize GTC/OSIRIS optical spectrum of GCS0845 (black
  thick line), classified as a L dwarf member candidate
  of Praesepe, based on proper motion and photometry. Overplotted are
  spectra of field dwarfs \citep{kirkpatrick1999} observed with
  the same instrumental configuration: 2M0251$+$2521 (M9; red dotted
  line), 2M0345$+$2540 (L0; dash-dotted line), and 2M0030$+$3139 (L2;
  red dash line).  All spectra are normalised at 7500\AA{} with offset
  of $+1$ between each spectra for clarity.}
\label{spectra-field}
\end{figure}

Among our targets, we confirmed spectroscopically the first L
dwarf in Praesepe (Fig.\,\ref{spectra-field}). We derive a spectral
type of L0.3$\pm$0.4, an effective temperature of 2279$\pm$371\,K, and
a mass of 71.1$\pm$23.0 M$_{\rm Jup}$, placing it at the
hydrogen-burning boundary.  The equivalent width of the gravity-sensitive
Na{\small{I}} doublet adds credit to the membership. We also derived a
probability of 79.5\% of being a member of Praesepe and argue that it
is a possible binary because of its location in various colour-magnitude 
diagrams \citep{boudreault2013}.

% FIGURE 3
%

\begin{figure}[t!]
\resizebox{\hsize}{!}{\includegraphics[clip=true]{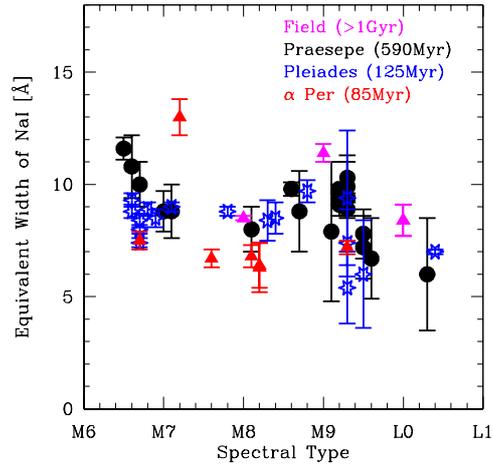}}
\caption{\footnotesize Equivalent widths of the Na{\small{I}} doublet
  as a function of spectral type for cluster member candidates in the
  Pleiades, $\alpha$~Per, Praesepe as well as field dwarfs.}
\label{na-spec}
\end{figure}

The NaI doublet at 8182/8194\AA{} is not resolved at our resolution.
The equivalent widths (EW) of the unresolved Na{\small{I}} doublet at
8188\AA{} are plotted as a function of spectral type in
Fig.\,\ref{na-spec} for the several member candidates in the three
clusters under study ($\alpha$~Per, the Pleiades, and Praesepe) and
for field M and L dwarfs observed with the same instrumental
configuration. For a given spectral type, the EWs increase with older
ages and decrease with cooler temperatures. The distribution of EWs
for candidates in Praesepe and the Pleiades suggests that Praesepe is
indeed older than the Pleiades.
 
% FIGURE 4
%

\begin{acknowledgements}
  Funding was provided by the Spanish ministry of science and
  innovation (program AYA2010-19136; PI is NL). NL is a Ram\'on
  y Cajal fellow at the IAC (program number 08-303-01-02).
\end{acknowledgements}

\bibliographystyle{aa}

\end{document}